\documentclass[prl,twocolumn,floatfix, nofootinbib,preprintnumbers,longbibliography]{revtex4-2}
\usepackage{graphicx}  
\usepackage{dcolumn}   
\usepackage{bm,relsize}        
\usepackage{amssymb, amsmath, amsfonts}
\usepackage{textcomp}
\usepackage{wasysym}
\usepackage{slashed}
\usepackage[caption=false]{subfig}
\usepackage{url}
\usepackage{bm}
\usepackage{braket}
\usepackage{bbold}
\usepackage{multirow}

\usepackage{lipsum, color}
\usepackage[usenames,dvipsnames,svgnames]{xcolor}
\usepackage{hyperref}
\hypersetup{colorlinks, citecolor=blue, linkcolor=blue, urlcolor=blue}
\usepackage{dsfont}
\usepackage[utf8]{inputenc} 

\usepackage{multirow} 
\allowdisplaybreaks

\begin{document}

\title{Time-Dependent Wave-Packet Approach to Reactions for Quantum Computation}

\author{Evan Rule}
\affiliation{Theoretical Division, Los Alamos National Laboratory, Los Alamos, NM 87545, USA}

\author{Ionel Stetcu}
\affiliation{Theoretical Division, Los Alamos National Laboratory, Los Alamos, NM 87545, USA}

\date{\today}
\preprint{LA-UR-26-22422}

\begin{abstract}
    We describe a method for obtaining the scattering matrix for nuclear or chemical reactions on a finite lattice. Aside from the preparation of the initial and final states as wave packets, the only other operation required is unitary time evolution, making this approach ideal for simulations on quantum hardware. The central quantity is a time-dependent overlap between incoming and outgoing wave packets whose Fourier transform corresponds to the scattering matrix at fixed energy, from which one can calculate elastic and inelastic cross sections for reactions involving two interacting clusters. Working in Cartesian coordinates enables an efficient encoding of the problem on quantum hardware via the first quantization mapping, with favorable qubit scaling for describing asymptotic scattering states. Within this framework, we describe a quantum algorithm for probing the scattering amplitude through different angles, including the forward direction, which provides access to the total cross section via the optical theorem. We demonstrate our methods through a series of numerical examples, for both elastic and inelastic processes, comparing against exact calculations. The techniques we describe can more readily be extended to a large number of constituent particles than other existing approaches, once fault-tolerant quantum hardware becomes available.
\end{abstract}

\maketitle
Quantum scattering theory is governed by the $S$ matrix, which encodes the transition amplitude between asymptotic states \cite{Taylor:1972pty}. The $S$ matrix is sufficiently general to describe complex reaction processes, including the breakup of a composite system into many possible final-state particles/channels. Nuclei composed of many individual nucleons have been successfully described by discretization to a finite lattice \cite{Lee:2008fa}. At low energies with relatively few open channels, it is possible to extract scattering phase shifts from a lattice whose spatial extent is large enough that the scattering wave function achieves its asymptotic form \cite{PhysRevLett.99.022502,PhysRevC.36.27,Carlson:2014vla}. The size of the many-body basis, however, grows rapidly as more particles are introduced, especially for particles carrying spin, isospin, or other quantum numbers. Without additional simplifying assumptions and/or phenomenological parameters (e.g., $R$-matrix theory \cite{RevModPhys.30.257,CARLSON198447,PhysRevC.79.044606,PhysRevLett.101.092501,ThompsonNunes2009,Descouvemont:2010cx,QUAGLIONI2025123095}), the problem quickly becomes intractable on classical computers. Fortunately, such systems can be efficiently encoded on a quantum computer in the first quantization mapping \cite{Abrams:1997gk,Berry:2018ggo,Weiss:2024mie,10.1063/5.0239980,*10.1063/5.0258298,Spagnoli:2025xvk}. 

Several algorithms have been proposed to calculate elastic scattering phase shifts on a quantum computer \cite{Sharma:2023bqu,Turro:2024ksf,Wang:2024scd,Yusf:2024igb}. To reliably extract cross sections with such methods, one must compute the phase shift in partial waves up to $\ell_\mathrm{max}\approx kR_0$, where $k$ is the scattering momentum and $R_0$ is the range of the potential. Moreover, these approaches are not straightforward to generalize to describe reactions with different initial- and final-state particles (i.e., capture or breakup reactions) or scattering between composite systems with internal structure that allows for inelastic transitions.

In this paper, we describe a method --- based on the time-dependent scattering formalism of Tannor \& Weeks \cite{10.1063/1.464016} --- for computing total and differential scattering cross sections using wave packets localized within a discrete lattice volume. Time-dependent wave-packet methods have long been employed in classical calculations of chemical reaction dynamics (see, e.g., \cite{NEUHAUSER1991460,annurev.pc.46.100195.001343,Chu01012006} and references therein). As the wave packets have a finite region of support, they can be propagated to asymptotic separation while remaining localized within the computational volume. Each wave packet contains a mixture of different momentum eigenstates; as a result, a single simulation provides information about the reaction process across a range of scattering energies. Via Fourier transform, we can isolate particular energy components of the $S$ matrix, from which the desired cross sections can be readily obtained.

 Our reaction formalism is directly compatible with a first-quantized, lattice-based description of the scattering constituents on quantum hardware: To each particle, we associate some number of qubits that encode the position of the particle within the spatial (or momentum) lattice as well as any additional quantum numbers (e.g., spin and/or isospin). Within this framework, we describe quantum algorithms to obtain total and differential cross sections for $2\rightarrow 2$ scattering, including inelastic processes where the scattering constituents transition between internal eigenstates. A schematic overview of our formalism is presented in \cite{supp}. 
 
 In the present work, we restrict our attention to short-ranged interactions characteristic of the strong nuclear interaction, but the necessary extension of the $S$-matrix scattering formalism to include Coulomb interactions exists and will be important to implement so that one can simulate charged-particle reactions.

We assume that the initial state is composed of two systems (i.e., a projectile and a target) with respective masses $M_1$, $M_2$, both of which in general may be composite. The total system is governed by the Hamiltonian
\begin{equation}
    H=H_0+H_\mathrm{int}=T_\mathrm{CM}+T_\mathrm{rel}+h^{(1)}+h^{(2)}+H_\mathrm{int},
\end{equation}
where $T_\mathrm{CM}$ and $T_\mathrm{rel}$ are kinetic energy operators corresponding, respectively, to the center-of-mass (CM) of the total system and the relative coordinate $\mathbf{r}=(x,y,z)$ that describes the separation of the two clusters; $h^{(i)}$ is the internal Hamiltonian of cluster $i$ consisting of the relative kinetic energy of the constituent particles and their interparticle (intracluster) interactions. The intercluster interaction $H_\mathrm{int}$ is assumed to be finite-range, so that the interaction vanishes (or is exponentially small) when the relative separation $r$ between the two clusters is sufficiently large. Each cluster's internal degrees of freedom are assumed to occupy an eigenstate $\ket{\psi^{(i)}_{\alpha_i}}$ satisfying $h^{(i)}\ket{\psi^{(i)}_{\alpha_i}}=E^{(i)}_{\alpha_i}\ket{\psi^{(i)}_{\alpha_i}}$, with $\alpha_i$ denoting a set of quantum numbers that completely specifies the state. The system will be prepared in a state with zero CM momentum and relative component described by a propagating Gaussian wave packet,
\begin{equation}
\begin{split}
    \phi_\mathrm{rel}(\mathbf{k}_0,\mathbf{r})=\frac{e^{-z^2/2\sigma^2}}{\sqrt{\sigma}\pi^{1/4}} \frac{e^{i\mathbf{k}_0\cdot\mathbf{r}}}{2\pi}, 
    \label{eq:phi_rel}
    \end{split}
\end{equation}
so that the total state of the system can be expressed as
\begin{equation}
    \ket{\Phi_{\mathbf{k}_0,\bm{\alpha}}}\equiv\ket{\psi_{\alpha_1}^{(1)}}\ket{\psi_{\alpha_2}^{(2)}}\phi_\mathrm{rel}(\mathbf{k}_0,\mathbf{r}),
\end{equation}
where $\mathbf{k}_0=(k_{0,x},k_{0,y},k_{0,z})$. Along the $\hat{\mathbf{z}}$ axis, the relative wave packet $\phi_\mathrm{rel}(\mathbf{k}_0,\mathbf{r})$ has spatial width $\sigma$ and propagates with group momentum $k_{0,z}$. Along the transverse directions, $\hat{\mathbf{x}}$ and $\hat{\mathbf{y}}$, the relative wave function occupies momentum eigenstates with respective eigenvalues $k_{0,x}$ and $k_{0,y}$.

The original proposal of Tannor \& Weeks \cite{10.1063/1.464016} is formulated in Jacobi coordinates, reducing the computational cost by eliminating the redundant CM degree of freedom. In contrast, our approach is based on absolute Cartesian coordinates, which map simply to a first-quantized encoding on quantum hardware. Moreover, in Ref. \cite{10.1063/1.464016}, momentum is imparted to the wave packet only along the direction of propagation. By introducing momentum $k_{0,x}$, $k_{0,y}$ along the transverse directions, we are able to probe scattering through different angles, as demonstrated below.

We are primarily interested in transitions between states of definite relative momentum,
\begin{equation}
\ket{\Psi_{\mathbf{k},\bm{\alpha}}}\equiv \ket{\psi^{(1)}_{\alpha_1}}\ket{\psi^{(2)}_{\alpha_2}}\sqrt{\frac{\mu}{k}}\frac{e^{i\mathbf{k}\cdot\mathbf{r}}}{(2\pi)^{3/2}},
\label{eq:free_scatter}
\end{equation} 
chosen here to be normalized with respect to energy. Such states are eigenstates of the non-interacting Hamiltonian $H_{0}$ with eigenvalue $E=k^2/2\mu+E_{\alpha_1}^{(1)}+E_{\alpha_2}^{(2)}$, where $\mu=M_1M_2/(M_1+M_2)$ is the reduced mass of the system. The notion of asymptotic states is made precise by introducing the M\o ller operators \cite{Taylor:1972pty}
\begin{equation}
    \Omega_{\pm}\equiv\lim_{t\rightarrow \mp\infty}e^{iHt}e^{-iH_0t}.
    \label{eq:moller_ops}
\end{equation}
When the M\o ller operators act on an eigenstate of the non-interacting Hamiltonian $H_{0}$, the resulting state $\ket{\Psi_{\mathbf{k},\bm{\alpha}}^{\pm}}\equiv\Omega_{\pm}\ket{\Psi_{\mathbf{k},\bm{\alpha}}}$ is an eigenstate of the fully interacting Hamiltonian $H$ with the same energy eigenvalue. Solutions obtained from the action of $\Omega_+$ ($\Omega_-$) are known as outgoing (incoming) scattering solutions.

Suppose that the final state consists of the same two clusters as the initial state; that is, the masses $M_1$, $M_2$ are conserved. (The kinematic modifications required to relax this assumption are straightforward.)  Choosing the Gaussian width $\sigma$ to be unchanged, we allow the initial and final wave packets to have different plane-wave factors, $\mathbf{k}_0=(k_{0,x},k_{0,y},k_{0,z})$ for the initial state and $\mathbf{k}_0'=(k_{0,x}',k_{0,y}',k_{0,z}')$ for the final state. The internal degrees of freedom of each cluster can transition between eigenstates of $h^{(i)}$, so that the initial and final states are labeled by quantum numbers $\bm{\alpha}=(\alpha_1,\alpha_2)$ and $\bm{\beta}=(\beta_1,\beta_2)$, respectively. We define the $S$ matrix to be the transition amplitude between outgoing and incoming scattering solutions
\begin{equation}
    \braket{\Psi_{\mathbf{k}',\bm{\beta}}^-|\Psi_{\mathbf{k},\bm{\alpha}}^+}=\delta(E'-E)S_{\bm{\beta},\bm{\alpha}}(E,\theta),
\end{equation}
where $E=k^2/2\mu+E_{\alpha_1}^{(1)}+E_{\alpha_2}^{(2)}$, $E'=k'^2/2\mu+E_{\beta_1}^{(1)}+E_{\beta_2}^{(2)}$, and $\cos\theta=\hat{\mathbf{k}}\cdot\hat{\mathbf{k}}'$ defines the scattering angle $\theta$. 

The free-particle solutions $\ket{\Psi_{\mathbf{k},\bm{\alpha}}}$ are plane waves in the relative coordinate $\mathbf{r}$ between the two clusters and thus are not spatially localized. The same is true of the corresponding scattering solutions $\ket{\Psi_{\mathbf{k},\bm{\alpha}}^{\pm}}$ --- beyond the range of the interaction, they coincide with the free solutions, $\Psi_{\mathbf{k},\bm{\alpha}}^\pm (\mathbf{r})\sim \Psi_{\mathbf{k},\bm{\alpha}}(\mathbf{r})$ as $r\rightarrow \infty$. As a result, it is computationally difficult to obtain the $S$ matrix directly from the scattering states $\ket{\Psi_{\mathbf{k},\bm{\alpha}}^\pm}$. Instead, we utilize the fact that the asymptotic scattering state can be obtained via energy projection of the corresponding M\o ller-evolved wave packet \cite{10.1063/1.464016,supp}, 
\begin{equation}
\begin{split}
        \ket{\Psi_{\mathbf{k},\bm{\alpha}}^{\pm}}&=\frac{k_z}{2\pi\mu \eta(k_z)}\sqrt{\frac{\mu}{k}}\int_{-\infty}^{+\infty}dt~e^{-iHt}\ket{\Phi_{\mathbf{k}_0,\bm{\alpha}}^{\pm}}e^{iEt},
\end{split}
\label{eq:energy_proj2}
\end{equation}
where $\ket{\Phi_{\mathbf{k}_0,\bm{\alpha}}^{\pm}}\equiv \Omega_{\pm }\ket{\Phi_{\mathbf{k}_0,\bm{\alpha}}}$, and the quantities $\mathbf{k}$, $\mathbf{k}_0$, and $E$ are related by
\begin{equation}
    k^2=k_{0,x}^2+k_{0,y}^2+k_z^2=2\mu\left(E-E^{(1)}_{\alpha_1}-E^{(2)}_{\alpha_2}\right).
    \label{eq:energy_condition}
\end{equation}
The normalization factor 
\begin{equation}
    \eta(k_z)=\frac{\sqrt{\sigma}}{\pi^{1/4}}e^{-(k_z-k_{0,z})^2\sigma^2/2},
    \label{eq:eta_rel}
\end{equation}
is the amplitude of the momentum mode $k_z$ within the Gaussian wave packet that governs the relative motion. Using Eq. \eqref{eq:energy_proj2}, we can express the $S$ matrix in terms of the M\o ller-evolved plane-wave states as \cite{10.1063/1.464016,supp} 
\begin{equation}
\begin{split}
    S_{\bm{\beta},\bm{\alpha}}(E,\theta)&=\frac{1}{2\pi\eta'(k_z')\eta(k_z)}\frac{k_zk_z'}{\mu\sqrt{kk'}}\\
    &\times\int_{-\infty}^{+\infty}dt~\bra{\Phi_{\mathbf{k}_0',\bm{\beta}}^-}e^{-iHt}\ket{\Phi_{\mathbf{k}_0,\bm{\alpha}}^+}e^{iEt},
    \end{split}
    \label{eq:elastic_s_matrix}
\end{equation}
where $\eta'(k_z')$ is analogous to $\eta(k_z)$ with $k_{z,0}\rightarrow k_{z,0}'$.  

The central quantity that can be obtained from quantum computation is the dimensionless overlap function
\begin{equation}
\mathcal{C}_{\bm{\beta},\bm{\alpha}}(\mathbf{k}_0',\mathbf{k}_0;t)\equiv\braket{\Phi_{\mathbf{k}_0',\bm{\beta}}^-|e^{-iHt}|\Phi_{\mathbf{k}_0,\bm{\alpha}}^+}.
        \label{eq:overlap}
\end{equation}
If the overlap function is sampled at sufficiently many times (set by the momentum/energy resolution of the lattice), the subsequent Fourier transform is trivial to perform classically. Assuming that the wave-packet states $\ket{\Phi_{\mathbf{k}_0,\bm{\alpha}}}$, $\ket{\Phi_{\mathbf{k}_0',\bm{\beta}}}$ can be efficiently prepared on a quantum computer, the only other required operation is unitary time evolution. On quantum hardware, the needed time evolution can be performed either by a product expansion (Trotterization) \cite{Trotter1959,*Suzuki:1985wzj} or via quantum signal processing (qubitization) \cite{Low:2016sck}. Reconstructing the $S$ matrix requires both the real and imaginary parts of the overlap function, obtained from respective Hadamard tests using a single set of particle qubit registers \cite{supp}. A Hadamard test requires $\mathcal{O}(\epsilon^{-2})$ measurements, where $\epsilon$ is the targeted error. The number of time samples will depend on the system and the simulation box size \cite{supp}.

The correlation function contains contributions to the $S$-matrix from all incident energies represented within the initial and final wave packets, and as a result the measured signal is large enough to be distinguished from the noise. Because the wave packets are extended plane waves in $\hat{\mathbf{x}}$ and $\hat{\mathbf{y}}$, the contributions to the cross section from all impact parameters are automatically included, further increasing the signal.

\begin{figure*}
    \includegraphics[scale=0.54]{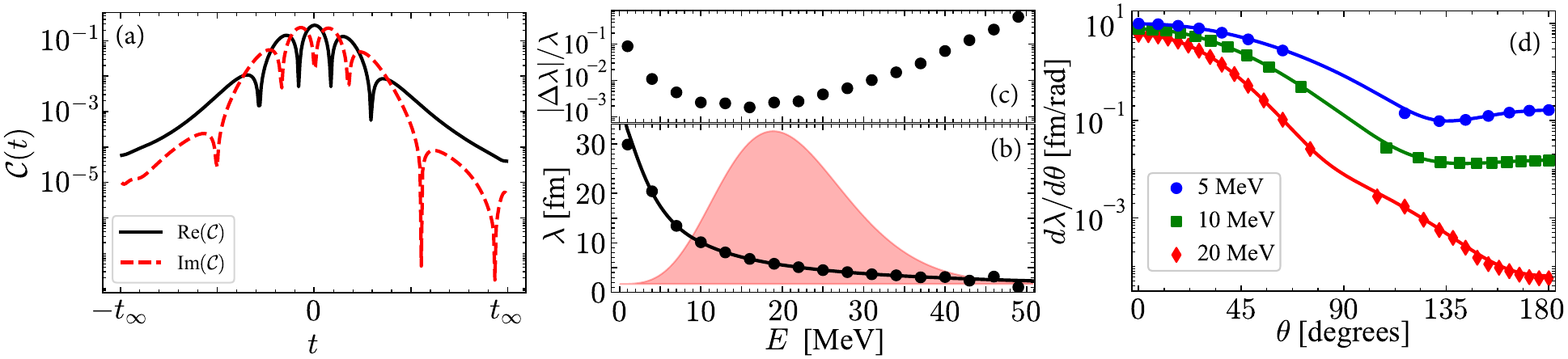}
    \caption{Elastic two-body scattering in two spatial dimensions by an attractive Gaussian potential [see Eq. \eqref{eq:test_potential}]. (a) Overlap function $\mathcal{C}(t)\equiv \mathcal{C}(\mathbf{k}_0',\mathbf{k}_0;t)$ for forward scattering, $\mathbf{k}_0'=\mathbf{k}_0=(0,k_{0,y})$. Solid black and dashed red lines denote, respectively, the real and imaginary parts of the overlap function. (b) Total cross section $\lambda$ as a function of energy, obtained from the forward-scattering amplitude via the optical theorem.  The red shaded region shows the relative energy profile of the wave packets (arbitrary normalization). (c) Relative error $|\Delta\lambda|/\lambda$ between the wave-packet and variable-phase results. (d) Differential cross section $d\lambda/d\theta$ at three different scattering energies, $E=5,10,$ and $20$ MeV. In panels (b) and (d), solid lines denote the exact numerical result obtained via the variable phase method \cite{alma991072475789706532,MARTINAZZO2003187} (adapted to two spatial dimensions); data points were obtained via the time-dependent wave-packet formalism.}
    \label{fig:2d_combined}
\end{figure*}

The procedure outlined above requires propagation of the initial and final wave packets to infinite times $t\rightarrow\pm \infty$, which is not feasible. However, it suffices to evolve the initial and final wave packets to an asymptotic time $t_\infty$, which is chosen to be large enough that the interaction between the wave packets is exponentially suppressed but small enough that they have not yet incurred significant edge effects from the periodic boundary conditions. As such, the $t\rightarrow \pm\infty$ limits appearing in the M\o ller operators, Eq. \eqref{eq:moller_ops}, can be replaced by a finite asymptotic time $\pm t_\infty$. Similarly, the limits of integration in Eq. \eqref{eq:elastic_s_matrix} can also be replaced by $\pm t_\infty$. The appropriate value of $t_\infty$ depends on the nature of the scattering problem, the wave packets, and the lattice size \cite{supp}. In principle, one can choose different asymptotic times $t_{-\infty}$ and $t_{+\infty}$ for $\Omega_+$ and $\Omega_-$ and the limits of integration of the overlap function. 

As discussed above, our formalism assumes a first-quantized description of the scattering constituents. In this way, the total kinetic energy operator $T=\sum_i T_i$  acts independently on each single-particle Hilbert space and therefore can be efficiently implemented on a quantum computer \cite{Rrapaj:2022hpd,Spagnoli:2025xvk}. The tradeoff is that either the initial or final wave packets must be projected onto a state with definite CM momentum, assumed here to be the zero-momentum state. As described further in \cite{supp}, the required projection can be efficiently performed with either standard quantum phase estimation (see e.g., \cite{PhysRevLett.125.230502}) or the projection algorithm introduced in Ref. \cite{Stetcu:2022nhy}. If we were to work directly in relative (or Jacobi) coordinates, as in Ref. \cite{10.1063/1.464016}, then the kinetic energy operator restricted to this subspace will not in general factorize across single-particle Hilbert spaces nor necessarily admit an efficient quantum circuit implementation.

The $S$ matrix obtained from a finite-volume calculation is a volume-dependent quantity --- for a fixed interaction, increasing the side length $L$ of the box decreases the proportion of the scattered wave relative to the incoming wave. Nonetheless, it is straightforward to exchange the volume-\textit{dependent} $S$ matrix for the volume-\textit{independent} scattering amplitude $f_{\bm{\beta},\bm{\alpha}}(E,\theta)$ \cite{1997qume.book.....M}, 
\begin{equation}
    S_{\bm{\beta},\bm{\alpha}}(E,\theta)=\delta_{\bm{\beta},\bm{\alpha}}\delta_{\theta}+\frac{2\pi i}{\sqrt{kk'}L^2}f_{\bm{\beta},\bm{\alpha}}(E,\theta),
    \label{eq:S_mat_to_f_amp}
\end{equation}
where $\delta_\theta=1$ if $\theta=0$ and $\delta_\theta=0$ if $\theta\neq 0$. The differential cross section $d\sigma_{\bm{\beta},\bm{\alpha}}/d\Omega$ and total cross section $\sigma_\mathrm{\bm{\beta,\bm{\alpha}}}$ corresponding to specific transitions between internal eigenstates $\bm{\alpha}\rightarrow\bm{\beta}$ can thus be obtained as
\begin{equation}
    \frac{d\sigma_{\bm{\beta},\bm{\alpha}}}{d\Omega}=|f_{\bm{\beta},\bm{\alpha}}(E,\theta)|^2,~~~~\sigma_{\bm{\beta},\bm{\alpha}}=\int d\Omega\frac{d\sigma_{\bm{\beta},\bm{\alpha}}}{d\Omega}.
\end{equation}

\textit{Numerical examples}. We demonstrate our methods with a series of numerical examples performed as classical computations. Due to the complexity of these calculations, we restrict our examples to two spatial dimensions. Scattering theory in two dimensions \cite{10.1119/1.14623} shares many of the pertinent features of its three-dimensional analog. We consider a system of two nucleons with $M_1=M_2=m_N=938.92$ MeV/$c^2$ interacting via the Gaussian potential
\begin{equation}
    V(|\mathbf{r}_1-\mathbf{r}_2|)=V_0U(r)=V_0\exp(-r^2/2R_0^2),
    \label{eq:test_potential}
\end{equation}
with $V_0 = -10$ MeV and $R_0=2.63$ fm. The resulting interaction is a satisfactory model of the strong nuclear force between two nucleons, resulting in a single bound state with energy $E_d=-2.225$ MeV, albeit in two spatial dimensions. Each particle has a Hilbert space corresponding to a $64^2$ spatial lattice with lattice constant $a=1.9$ fm. In the first quantization mapping, this system could be represented by 12 logical qubits per particle. The kinetic energy is treated exactly via Fourier transform; for implementation on quantum hardware, it may be advantageous to instead employ a finite difference approximation. The scattering constituents under consideration have no internal degrees of freedom, and we will hereafter omit the eigenstate labels $\bm{\alpha},\bm{\beta}$.

The wave packets were chosen to have spatial width $\sigma=5.39$ fm. The initial and final wave-packet momenta were chosen to be $\mathbf{k}_0=(0,k_{0,y})$, $\mathbf{k}_0'=(k_{0,x}',k_{0,y}')$ with $|k_{0,y}|=|k_{0,y}'|=0.644$ fm$^{-1}$. (In two spatial dimensions, we assume that $\hat{\mathbf{y}}$ is the direction along which the wave packet propagates.) To consider scattering through angles $0^\circ\leq \theta\leq 90^\circ$, we take $k_{0,y}=k_{0,y}'$ and vary the transverse momentum of the final state $k_{0,x}'$ over all allowed lattice momenta. Backscattering ($90^\circ\leq \theta\leq 180^\circ$) is assessed via the same procedure but with $k_{0,y}=-k_{0,y}'$. (In three spatial dimensions, the scattering angle $\cos\theta=\hat{\mathbf{k}}\cdot\hat{\mathbf{k}}'$ can be varied analogously, thus avoiding the cumbersome formulation of Jacobi coordinates.) To prepare the M\o ller states, the wave packets were evolved to time $t_\infty=1.08$ MeV$^{-1}$. Further technical details are provided in \cite{supp}. 

Forward scattering corresponds to $k_{0,x}'=0$. After forming the M\o ller-evolved wave packets $\ket{\Phi_{\mathbf{k}_0}^+}$, $\ket{\Phi_{\mathbf{k}_0'}^-}$ one obtains the overlap function shown in Fig. \ref{fig:2d_combined}(a). Performing the Fourier transform of this overlap function, as in Eq. \eqref{eq:elastic_s_matrix}, one can obtain the forward-scattering matrix $S(E,0)$ at any energy $E$ where the wave packets have significant support. Relating this result to the forward-scattering amplitude $f(E,0)$ via the two-dimensional analog of Eq. \eqref{eq:S_mat_to_f_amp} (see \cite{supp}), one can obtain the total cross section from the two-dimensional optical theorem \cite{10.1119/1.14623}, $\lambda=\sqrt{8\pi}/k~\mathrm{Im}f(E,0)$.

The considered energy need not correspond to the energy of one of the lattice eigenstates; although the lattice calculation consists of a discrete basis of particular eigenmodes, the resulting overlap function smoothly interpolates between them. This property is evident in Fig. \ref{fig:2d_combined}(b), where the total cross section is computed at evenly spaced scattering energies $E=1,4,...,49$ MeV, which do not correspond to lattice eigenenergies. As shown in the same figure, the chosen wave packets have appreciable support for scattering energies $5\lesssim E \lesssim 40$ MeV. Within this region, the relative error on the cross section is typically much less than $1\%$, as shown in Fig. \ref{fig:2d_combined}(c).

Figure \ref{fig:2d_combined}(d) shows the angular differential cross section $d\lambda/d\theta$ at three different scattering energies, $E=5$, $10$, and $20$ MeV. As the transverse momentum $k_{0,x}'$ must correspond to a lattice eigenstate, only a specific finite set of scattering angles can be probed at a given energy, with higher scattering energies admitting more possible angles. The agreement between the wave-packet solution and that obtained by numerical integration in partial waves is generally quite good, even when $\theta \approx \pi$ and the scattering is very weak. Close to $\theta=\pi/2$, the scattering is dominated by momentum modes $k_{0,x}'$ along the transverse direction; in such cases, the projected final-state momentum $k'_{y}$ is very close to zero and not well represented by the wave packet, leading to comparatively large errors in the cross section.

Existing quantum algorithms for elastic nuclear scattering \cite{Sharma:2023bqu,Turro:2024ksf,Wang:2024scd,Yusf:2024igb} allow one to calculate the partial-wave phase shifts. Depending on the nature of the interaction, constructing the corresponding cross sections could potentially require summation over many partial waves. In the case of a non-central potential, such as the nuclear tensor interaction, the partial-wave phase shifts can only be obtained by solving a system of coupled equations. Our approach is completely agnostic to the nature of the interaction, as long as it is short ranged. Though for convenience we consider here a central potential, our framework operates identically if the interaction permits the exchange of angular momentum (or other quantum numbers such as spin and isospin) between projectile and target.

\textit{Inelastic scattering}. To demonstrate the generality of our formalism, we extend the previous example to a two-channel scattering problem with Hamiltonian
\begin{equation}
    H=\left(\begin{array}{cc}
     T+V_0U(r)    & \bar{V}_0U(r) \\
       \bar{V}_0U(r)  & T+\Delta+2V_0U(r) 
    \end{array}\right),
\end{equation}
with $U(r)$ as in Eq. \eqref{eq:test_potential}, and $V_0=10$ MeV, $\bar{V}_0=14.24$ MeV, and $\Delta=13$ MeV, chosen such that the system does not have any Feshbach resonances. Let us label the zero-threshold channel by the quantum number $\alpha=0$ and the finite-threshold $(\Delta>0)$ channel by $\alpha= 1$. We otherwise adopt identical lattice and wave-packet parameters to those employed in the previous example.

Preparing both the initial and final wave packets in the zero-threshold channel, we obtain the elastic scattering matrix $S_{0,0}(E,\theta)$ and, correspondingly, the elastic differential cross section $d\lambda_{0,0}/d\theta$. Likewise, preparing the initial wave packet in the zero-threshold channel and the final wave packet in the $\Delta$-threshold channel, we obtain the inelastic scattering matrix $S_{1,0}(E,\theta)$ and the inelastic differential cross section $d\lambda_{1,0}/d\theta$. As before, we can probe scattering through different angles by imparting transverse momentum to the final-state wave packets. 

\begin{figure}
    \centering
    \includegraphics[scale=0.6]{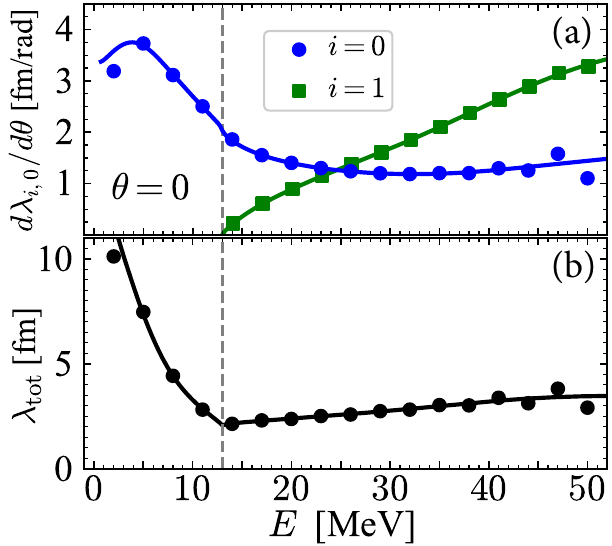}
    \caption{Two-channel scattering cross sections: (a) Elastic (blue) and inelastic (green) differential cross sections evaluated in the forward limit $\theta=0$. (b) Total (elastic + inelastic) integrated cross section $\lambda_\mathrm{tot}$ obtained via the optical theorem. Solid lines were calculated via the variable-phase method; points correspond to the time-dependent wave-packet formalism. Vertical dashed line denotes the excited state threshold energy, $\Delta=13$ MeV.}
    \label{fig:inelastic}
\end{figure}

Figure \ref{fig:inelastic}(a) shows the elastic and inelastic differential cross sections evaluated in the forward limit. As expected, the inelastic cross section vanishes at the threshold energy $\Delta=13$ MeV. Figure \ref{fig:inelastic}(b) shows the total integrated cross section $\lambda_\mathrm{tot}=\lambda_{0,0}+\lambda_{1,0}$ obtained from the forward-scattering amplitude $f_{0,0}(E,0)$ via the optical theorem. As in the previous example, the wave-packet formalism performs very well at energies where the wave packets have appreciable support. [The energy profile employed here is the same as that shown in Fig. \ref{fig:2d_combined}(b).]

\textit{Summary.} In this paper, we describe a time-dependent wave-packet approach to scattering that is ideal for quantum computation in the first quantization mapping. So far, we have restricted our attention to elastic and inelastic scattering of two particles interacting via a short-ranged potential. The approach, however, is general --- both the target and the projectile can be composite systems, opening the way for simulation of nuclear and chemical reactions. A detailed error analysis of our formalism is provided in the supplemental material \cite{supp}.

The use of wave packets localized to a finite lattice provides scattering information across the entire range of energies where the wave packets have appreciable support. Naturally, the wave packets and the lattice are chosen so that the two clusters can be brought to asymptotic separation without incurring edge effects. The most straightforward means of simulating scattering at lower energies is therefore to increase the spatial extent of the lattice. 

Interplay between the wave packets and the lattice also determines the time over which the system can be evolved and over which the central quantity --- the overlap function --- can be probed. For the relatively fast reactions considered here, the overlap function damps exponentially at long times. For scattering involving resonances or compound states, the overlap function is expected to exhibit long-lived behavior, which can be resolved by employing a lattice with a larger spatial extent. In first quantization, the side length $L$ of a cubic lattice is doubled by introducing one additional qubit per particle per spatial dimension. 

In realistic calculations involving identical particles, the scattering constituents must have the proper exchange symmetry (i.e., symmetric exchange for bosons, antisymmetric for fermions), which is not guaranteed a priori of many-body states in first quantization. Fully antisymmetric states can be prepared using one of several antisymmetrization algorithms \cite{Berry:2018ggo,10.1063/5.0239980,*10.1063/5.0258298,Rule:2025ldr}, including one specialized for antisymmetrizing the constituent particles of a target and projectile that are spatially separated \cite{Stetcu:2025bbu}. 

Elastic/inelastic cross sections can then be calculated by preparing the outgoing products into properly antisymmetrized ground/excited states. While our method targets a specific final state, it naturally accounts for scattering into all energetically open channels. Extending this framework to treat reactions with more than two final-state clusters will require significant modification; the final-state kinematics are less constrained than the two-cluster case, allowing for different momentum-sharing between the outgoing clusters.  More straightforward variations of the method outlined in this work will allow for the simulation of charged-particle reactions where the Coulomb force acts between the scattering constituents, as well as the treatment of photo-induced reactions. 

\section*{Acknowledgments}
The authors are grateful to Joe Carlson and Ronen Weiss for helpful discussions. This work was carried out under the auspices of the National Nuclear Security Administration of the U.S. Department of Energy at Los Alamos National Laboratory under Contract No. 89233218CNA000001. The research presented in this article was partially supported by the Laboratory Directed Research and Development program of Los Alamos National Laboratory under project numbers 20251163PRD3 and 20260043DR. ER was supported by the National Science Foundation under cooperative agreement 2020275 during the partial completion of this work. IS acknowledges partial support from the Advanced Simulation and Computing (ASC) program. 

\bibliography{QScatter}

\end{document}


\title{Supplemental material for: Time-Dependent Wave-Packet Approach to Reactions for Quantum Computation}

\author{Evan Rule}
\affiliation{Theoretical Division, Los Alamos National Laboratory, Los Alamos, NM 87545, USA}

\author{Ionel Stetcu}
\affiliation{Theoretical Division, Los Alamos National Laboratory, Los Alamos, NM 87545, USA}

\date{\today}

\maketitle

\section{Algorithmic Overview}
\begin{figure}
    \centering
    \includegraphics[scale=1.0]{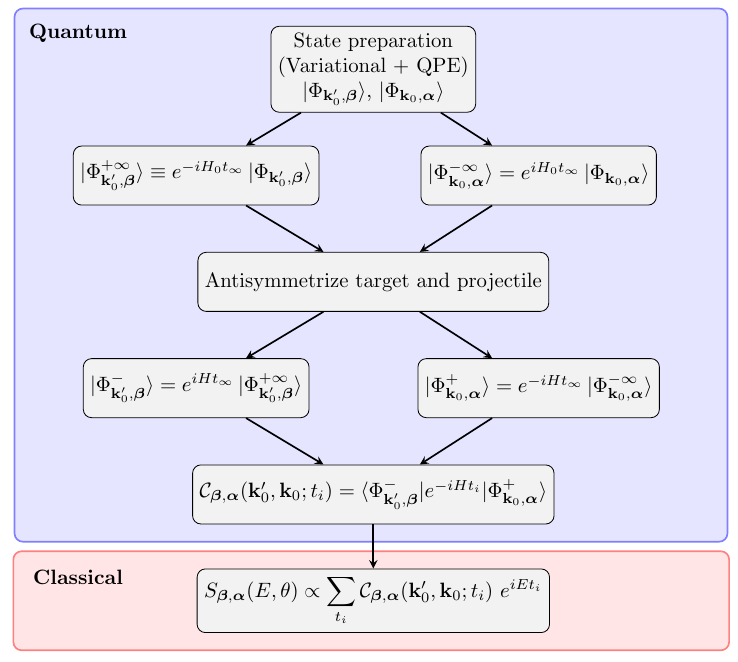}
    \caption{Schematic overview of the proposed quantum reaction algorithm.}
    \label{fig:algorithm}
\end{figure}

Figure \ref{fig:algorithm} provides an illustration of our proposed framework: We begin by preparing the initial and final wave-packet states $\ket{\Phi_{\mathbf{k}_0,\bm{\alpha}}}$ and $\ket{\Phi_{\mathbf{k}_0',\bm{\beta}}}$. Separate techniques are required to prepare the internal eigenstates $\ket{\psi^{(i)}_{\alpha_i}}$ and the relative wave packet $\phi_\mathrm{rel}(\mathbf{k}_0,\mathbf{r})$. The latter is discussed below in Sec. \ref{app:CM_rel}. The eigenstates can be prepared, for example, starting from a totally antisymmetric trial wave function for each cluster (e.g., Hartee-Fock); if necessary, this trial wave function can be improved through variational techniques before projection to the desired energy eigenstate using quantum phase estimation (QPE). Note that, at this point, the system is totally antisymmetric under exchanges of particles within each cluster, but not across the clusters. 

Next, we perform the first half of the time evolution required to prepare the M\o ller states; that is, the initial (final) wave packet is evolved with the non-interacting Hamiltonian $H_0$ to the asymptotic past (future). At this point, the two clusters are well separated within the computational volume. This fact can be exploited in order to efficiently perform the final antisymmetrization between particles in the projectile with those in the target \cite{Stetcu:2025bbu}. Once this has been achieved, the preparation of the M\o ller states is completed by evolving with the fully interacting Hamiltonian $H$.

The final step is the extraction of the overlap function whose Fourier transform corresponds to the scattering matrix. In the simplest formulation, we discretize the time interval from $[-t_\infty,t_\infty]$ into $N$ evenly spaced timesteps $t_i$. For each $t_i$, we evolve the initial M\o ller state with $e^{-iHt_i}$ and then extract the overlap function $\mathcal{C}_{\bm{\beta},\bm{\alpha}}(\mathbf{k}_0',\mathbf{k}_0,t_i)$, as detailed in Sec. \ref{app:overlap_quantum} below. Once collected, the overlap function samples are post-processed classically to produce the $S$ matrix.

\section{Separation of relative and center-of-mass components}
\label{app:CM_rel}
As in the main text, assume that the initial state is composed of two systems (i.e., a projectile and a target) with respective masses $M_1$, $M_2$, both of which in general may be composite. The location of each system is described by a center of mass (CM) coordinate 
\begin{equation}
    \mathbf{R}_i=(X_i,Y_i,Z_i)=\frac{\sum_{j}m_{i,j}~\mathbf{r}_{i,j}}{M_i},~~~~~i=1,2
\end{equation}
where $\mathbf{r}_{i,j}$ is the coordinate of constituent particle $j$ within cluster $i$ and $m_{i,j}$ is the corresponding mass. The two clusters are governed by respective Hamiltonians
\begin{equation}
\begin{split}
    H^{(i)}&=T^{(i)}_\mathrm{CM}+h^{(i)},~~~~~i=1,2
\end{split}
\end{equation}
where 
\begin{equation}
    T^{(i)}_\mathrm{CM}=\frac{\mathbf{P}_i^2}{2M_i}=\frac{1}{2M_i}\left(\sum_j \mathbf{p}_{i,j}\right)^2,
\end{equation}
is the kinetic energy operator corresponding to CM coordinate $\mathbf{R}_i$, and $h^{(i)}$ is the internal Hamiltonian of cluster $i$ consisting of the relative kinetic energy of the constituent particles and their interparticle (intracluster) interactions. Thus the total Hilbert space can be factorized as 
\begin{equation}
    \mathcal{H}=\mathcal{H}^{(1)}\otimes\mathcal{H}^{(2)}=\mathcal{H}^{(1)}_\mathrm{rel}\otimes \mathcal{H}^{(1)}_\mathrm{CM}\otimes\mathcal{H}^{(2)}_\mathrm{rel}\otimes\mathcal{H}^{(2)}_\mathrm{CM},
\end{equation}
where $T^{(i)}_\mathrm{CM}$ acts on $\mathcal{H}^{(i)}_\mathrm{CM}$ and $h^{(i)}$ acts on $\mathcal{H}^{(i)}_\mathrm{rel}$. The relative and center-of-mass coordinates for the total system (projectile + target) are then given by
\begin{equation}
\begin{split}
    \mathbf{r}&=\mathbf{R}_1-\mathbf{R}_2=(x,y,z),\\
    M\mathbf{R}&=M_1\mathbf{R}_1+M_2\mathbf{R}_2=M(X,Y,Z),
\end{split}
\end{equation}
where $M=M_1+M_2$ is the total mass. Thus we additionally factorize
\begin{equation}
    \mathcal{H}^{(1)}_\mathrm{CM}\otimes \mathcal{H}^{(2)}_\mathrm{CM}=\mathcal{H}_\mathrm{rel}\otimes\mathcal{H}_\mathrm{CM}.
\end{equation}

In the main text, we assumed that the total system would be prepared directly in the state 
\begin{equation}
    \ket{\Phi_{\mathbf{k}_0,\bm{\alpha}}}\equiv\ket{\psi_{\alpha_1}^{(1)}}\ket{\psi_{\alpha_2}^{(2)}}\phi_\mathrm{rel}(\mathbf{k}_0,\mathbf{r}),
    \label{eq:Phi_rel}
\end{equation}
with
\begin{equation}
    \phi_\mathrm{rel}(\mathbf{k}_0,\mathbf{r})=\frac{e^{-z^2/2\sigma^2}}{\sqrt{\sigma}\pi^{1/4}} \frac{e^{i\mathbf{k}_0\cdot\mathbf{r}}}{2\pi}, 
    \label{eq:wp_def}
\end{equation}
where $\mathbf{k}_0=(k_{0,x},k_{0,y},k_{0,z})$ specifies the momentum of the plane-wave factor. Recall that $\ket{\psi^{(i)}_{\alpha_i}}$ is an eigenstate of the internal cluster Hamiltonian $h^{(i)}$. Directly preparing the state in Eq. \eqref{eq:Phi_rel} in a first-quantized Hilbert space would require quantum operations that act simultaneously on all qubits associated to the constituent particles of both clusters. Alternately, we can prepare each cluster separately in the state
\begin{equation}
    \ket{\Phi^{(i)}_{\mathbf{k}_0,\alpha_i}}=\ket{\psi^{(i)}_{\alpha_i}}\phi_i(\mathbf{k}_0,\mathbf{R}_i),
    \label{eq:cluster_state}
\end{equation}
where
\begin{equation}
\begin{split}    
    \phi_1(\mathbf{k}_0,\mathbf{R}_1)&=\frac{e^{-Z_1^2/2\sigma_1^2}}{\sqrt{\sigma_1}\pi^{1/4}}\frac{e^{i\mathbf{k}_0\cdot\mathbf{R}_1}}{2\pi}, \\
    \phi_2(\mathbf{k}_0,\mathbf{R}_2)&=\frac{e^{-Z_2^2/2\sigma_2^2}}{\sqrt{\sigma_2}\pi^{1/4}}\frac{e^{- i\mathbf{k}_0\cdot\mathbf{R}_2}}{2\pi}.
    \end{split}
    \label{eq:wp_def}
\end{equation}
The state preparation is then factorized between the respective Hilbert spaces of the two clusters, significantly reducing the complexity of the required quantum operations. 

We will now briefly outline a method to prepare each cluster in a state of the form in Eq. \eqref{eq:cluster_state}. In principle, the CM can be localized within a Gaussian wave packet by adding to the physical Hamiltonian an additional harmonic oscillator potential for the CM degree of freedom, $V_\mathrm{CM}(\mathbf{R}_i)=\frac{1}{2}M_i\omega^2 \mathbf{R}^2_i$. This harmonic trap introduces terms of the form $\mathbf{r}_{i,j}\cdot\mathbf{r}_{i,k}$ for all particle pairs $j\neq k$ in cluster $i$, which scale exponentially in first quantization. Other ancillae-based alternatives that avoid this exponential scaling are being developed and will be the subject of future publication (S. Masot-Llima et al., unpublished).

If the cluster wave packets are chosen such that $\sigma_1/\sigma_2=\sqrt{M_2/M_1}$, then the center-of-mass and relative coordinates decouple so that the total state of the system is  
\begin{equation}
    \ket{\tilde{\Phi}_{\mathbf{k}_0,\bm{\alpha}}}\equiv\ket{\psi_{\alpha_1}^{(1)}}\ket{\psi_{\alpha_2}^{(2)}}\phi_\mathrm{rel}(\mathbf{k}_0,\mathbf{r})\phi_\mathrm{CM}(Z),
    \label{eq:wfn_with_cm}
\end{equation}
where
\begin{equation}
    \phi_\mathrm{CM}(Z)=\frac{1}{\sqrt{\Sigma}\pi^{1/4}}e^{-Z^2/2\Sigma^2}\frac{1}{2\pi}.
\end{equation}
The respective widths of the CM and relative wave packets satisfy $\Sigma= \sqrt{M_2/M}\sigma_2$, $\sigma= \sqrt{M/M_1}\sigma_2$.

We distinguish $\ket{\tilde{\Phi}_{\mathbf{k}_0,\bm{\alpha}}}$ from $\ket{\Phi_{\mathbf{k}_0,\bm{\alpha}}}$ based on the fact that the former contains a Gaussian wave packet for the CM whereas the latter has the CM in the zero-momentum eigenstate. The energy projector [Eq. (7) in the main text] that isolates particular components of the wave packet is based on unitary time evolution with the total Hamiltonian $H$; as a result, both relative and CM momentum modes contribute to the target energy $E$. Therefore, in order to isolate a particular relative momentum eigenstate $\ket{\Phi_{\mathbf{k},\bm{\alpha}}}$ from within the wave packet $\ket{\Phi_{\mathbf{k}_0,\bm{\alpha}}}$, we must ensure that the CM momentum of the wave packet is zero and thus makes no contribution to the projected energy $E$. To achieve this, we introduce the projection operator $\hat{P}_{\mathrm{CM}0}$ that isolates the zero-CM-momentum component of the wave packet,
\begin{equation}
    \hat{P}_{\mathrm{CM}0}\ket{\tilde{\Phi}_{\mathbf{k}_0,\bm{\alpha}}}=\frac{\sqrt{\Sigma}}{\pi^{1/4}}\ket{\Phi_{\mathbf{k}_0,\bm{\alpha}}}.
    \label{eq:CM_proj_op}
\end{equation}
Even with the additional projection, this method of separately preparing the cluster wave packets should be more efficient than directly preparing $\ket{\Phi_{\mathbf{k}_0,\bm{\alpha}}}$. In practice, it is only necessary to project either the initial or final wave packet onto the zero CM eigenstate. Once prepared, $\ket{\Phi_{\mathbf{k}_0,\bm{\alpha}}}$ is acted on with unitary time evolution with either $H_0$ or $H$, both of which will preserve the CM of the total system (assuming that the interactions are invariant under translations of the total CM of the system). For example, if the initial state (but not the final state) is prepared with zero CM momentum, then the overlap function can be computed as
\begin{equation}
    \begin{split}
        \braket{\tilde{\Phi}^-_{\mathbf{k}_0',\bm{\beta}}|e^{-iHt}|\Phi^+_{\mathbf{k}_0,\bm{\alpha}}}&=\braket{\Phi^-_{\mathbf{k}_0',\bm{\beta}}|e^{-iHt}|\Phi^+_{\mathbf{k}_0,\bm{\alpha}}}\int dZ~\phi_\mathrm{CM}(Z) \\
        &=\sqrt{2\Sigma}\pi^{1/4}\mathcal{C}_{\bm{\beta},\bm{\alpha}}(\mathbf{k}_0',\mathbf{k_0};t).
        \label{eq:CM_overlap_function}
    \end{split}
\end{equation}

As evident from Eq. \eqref{eq:wfn_with_cm}, the CM momentum is directed along the $\hat{\mathbf{z}}$ axis. Above we have assumed that we will project onto the state with CM momentum equal to zero. In actuality, our scattering formalism can be applied whenever the CM component is projected into any definite momentum eigenstate, not necessarily $K_z=0$. In this way, the only undetermined component of the wave-packet system is the relative momentum along the $\hat{\mathbf{z}}$ direction, which remains in a Gaussian profile until it is projected into a definite momentum eigenstate via Eq. (7) in the main text. After projection, the total energy of the two-cluster system is
\begin{equation}
    E=E_{\alpha_1}^{(1)}+E_{\alpha_2}^{(2)}+k^2/2\mu+K_z^2/2M.
    \label{eq:energy_shift_CM}
\end{equation}
To achieve the desired CM projection, we can utilize either standard QPE or the projection method described in Ref. \cite{Stetcu:2022nhy}. In either case, one performs unitary time evolution by exponentiating the operator $\hat{K}_z=\sum_{i}\hat{k}_{z,i}$, which acts independently on each single-particle Hilbert space and only on those qubits that encode the $z$ coordinate of a given particle. Implementing QPE with initial state $\ket{\tilde{\Phi}_{\mathbf{k}_0,\bm{\alpha}}}$ and unitary operator $U=e^{i\hat{K}_z}$ projects onto some eigenstate of the operator $\hat{K}_z$, not necessarily $K_z=0$. As the CM momentum distribution is peaked at $K_z=0$, this is the most probable outcome, but if any other value of $K_z$ is measured, we can correct the resulting correlation function by properly accounting for the CM contribution to the total energy, as in Eq. \eqref{eq:energy_shift_CM}. One must also make the obvious modifications to Eqs. \eqref{eq:CM_proj_op} and \eqref{eq:CM_overlap_function} to properly account for the amplitude of momentum mode $K_z$ within the CM wave packet. Thus we can recover the $S$ matrix for a specific relative momentum transfer $\mathbf{k'}-\mathbf{k}$, even when the correlation function samples were obtained from many different values of the CM momenta. 

Using the method of Ref. \cite{Stetcu:2022nhy}, one attempts to project onto the $K_z=0$ component with success probability given by the overlap of the initial CM wave packet with the $K_z=0$ component. In case of failure, the system is projected into the orthogonal component, and one can repeat this procedure with a different targeted CM momentum $K_z$, based on all previous measurements, until success. Compared to standard QPE, this method requires only a single ancilla qubit that, in case of failure, can be reused in subsequent attempts. We are currently investigating another projection approach that avoids time evolution by using a set of ancillae correlated with the physical system (S. Masot-Llima et al., unpublished).

\section{Details on the derivation of Eqs. (7) and (10)}
\label{app:derivation}
Expressing the relative wave packet in terms of its Fourier components as
\begin{equation}
    \phi_\mathrm{rel}(\mathbf{k}_0,\mathbf{r})=\frac{1}{2\pi}e^{ik_{0,x}x+ik_{0,y}y}\int_{-\infty}^{+\infty} dk_z~\eta(k_z)\frac{e^{ik_zz}}{\sqrt{2\pi}},
    \label{eq:fourier_decomp}
\end{equation}
with 
\begin{equation}
    \eta(k_z)=\frac{\sqrt{\sigma}}{\pi^{1/4}}e^{-(k_z-k_{0,z})^2\sigma^2/2},
    \label{eq:eta_rel}
\end{equation}
we will now demonstrate how a specific relative-momentum eigenstate $\mathbf{k}=(k_{0,x},k_{0,y},k_z)$ can be isolated via time evolution with the non-interacting Hamiltonian $H_{0}=T_\mathrm{CM}+T_\mathrm{rel}+h_0^{(1)}+h_0^{(2)}$. The derivation proceeds as \footnote{Equation \eqref{eq:EQ10_deriv} contains a slight abuse of notation, as the momentum $k_z$ is first treated as an integration variable and later, after integration against the Dirac delta function, as a fixed quantity. This choice was made in order to simplify the notation, at the risk of possible confusion. The same abuse of notation is made in Eq. \eqref{eq:phi_psi_overlap} below.}
\begin{equation}
    \begin{split}    
      &\int_{-\infty}^{+\infty}dt~e^{-iH_{0}t}\ket{\Phi_{\mathbf{k}_0,\bm{\alpha}}}e^{iEt}\\
        &= \frac{1}{2\pi}e^{ik_{0,x}x+ik_{0,y}y}\ket{\psi^{(1)}_{\alpha_1}}\ket{\psi^{(2)}_{\alpha_2}}\int~dk_z~\eta(k_z)\frac{e^{ik_zz}}{\sqrt{2\pi}}\int_{-\infty}^{+\infty}dt~\exp\left[i\left(E-E^{(1)}_{\alpha_1}-E^{(2)}_{\alpha_2}-k^2/2\mu\right)t\right] \\
        &= 2\pi\frac{1}{2\pi}e^{ik_{0,x}x+ik_{0,y}y}\ket{\psi^{(1)}_{\alpha_1}}\ket{\psi^{(2)}_{\alpha_2}}\int dk_z~\eta(k_z)\frac{e^{ik_zz}}{\sqrt{2\pi}}~\delta\left(E-E^{(1)}_{\alpha_1}-E^{(2)}_{\alpha_2}-k^2/2\mu\right) \\
        &=2\pi\frac{1}{2\pi}e^{ik_{0,x}x+ik_{0,y}y}\frac{\mu}{k_z}\left[\eta(-k_z)\frac{e^{-ik_zz}}{\sqrt{2\pi}}+\eta(k_z)\frac{e^{ik_zz}}{\sqrt{2\pi}}\right]\ket{\psi^{(1)}_{\alpha_1}}\ket{\psi^{(2)}_{\alpha_2}},
    \end{split}
    \label{eq:EQ10_deriv}
\end{equation}
where $k_z$ is chosen to correspond to the positive square root; that is,
\begin{equation}
    k_z=+\sqrt{2\mu\left(E-E^{(1)}_{\alpha_1}-E^{(2)}_{\alpha_2}\right)-k_{0,x}^2-k_{0,y}^2}.
\end{equation}
Thus, the Fourier transform projects onto total energy $E$, which can, in principle, correspond to either right-moving ($e^{+ik_z z}$) or left-moving ($e^{-ik_zz}$) momentum along the relative $z$ coordinate. As shown in Eq. \eqref{eq:eta_rel}, the initial wave packet in the relative coordinate is a Gaussian with group momentum $k_{0,z}$. In order for the two clusters to achieve asymptotic separation, the zero-momentum component of their relative coordinate must be exponentially suppressed. That is, $k_{0,z}$ and $\sigma$ must be such that 
\begin{equation}
    \eta(0)=\frac{\sqrt{\sigma}}{\pi^{1/4}}e^{-k_{0,z}^2\sigma^2/2}\approx 0.
\end{equation}
If this is the case, then the left-moving components are also exponentially suppressed,
\begin{equation}
    \eta(-k_z)=\frac{\sqrt{\sigma}}{\pi^{1/4}}e^{-(k_z+k_{0,z})^2\sigma^2/2}\approx 0,
\end{equation}
where we have assumed that $k_{0,z}>0$. Hence, we neglect the left-moving component in Eq. \eqref{eq:EQ10_deriv}.

The M\o ller operators satisfy the intertwining relations \cite{Taylor:1972pty}
\begin{equation}
    H\Omega_{\pm}=\Omega_{\pm}H_{0},
    \label{eq:intertwining}
\end{equation}
allowing us to write
\begin{equation}
\begin{split}
     \Omega_+\int_{-\infty}^{+\infty}dt~e^{-iH_{0}t}\ket{\Phi_{\mathbf{k}_0,\bm{\alpha}}}e^{iEt}&=\int_{-\infty}^{+\infty}dt~e^{-iHt}\Omega_+\ket{\Phi_{\mathbf{k}_0,\bm{\alpha}}}e^{iEt}\\
     &= 2\pi\frac{\mu}{k_z}\sqrt{\frac{k}{\mu}}\eta(k_z)\Omega_+\left(\sqrt{\frac{\mu}{k}}\frac{e^{i\mathbf{k}\cdot\mathbf{r}}}{(2\pi)^{3/2}}\ket{\psi^{(1)}_{\alpha_1}}\ket{\psi^{(2)}_{\alpha_2}}\right)\\
     &=2\pi\frac{\mu}{k_z}\sqrt{\frac{k}{\mu}}\eta(k_z) \ket{\Psi_{\mathbf{k},\bm{\alpha}}^+},
     \end{split}
\end{equation}
which is equivalent to Eq. (7) in the main text. To proceed to Eq. (10), we expand the final-state wave packet as
\begin{equation}
    \ket{\Phi^-_{\mathbf{k}_0',\bm{\beta}}}=\int dk_{z}'~\eta'(k_z')\sqrt{\frac{k'}{\mu}}\ket{\Psi_{\mathbf{k}',\bm{\beta}}^-},
\end{equation}
where $\mathbf{k}'_0=\left(k_{0,x}',k_{0,y}',k_{0,z}'\right)$, $\mathbf{k}'=\left(k_{0,x}',k_{0,y'},k_z'\right)$ and $\eta'(k_z')$ is equivalent to $\eta(k_z)$ with $k_{0,z}\rightarrow k_{0,z}'$. This allows us to write
\begin{equation}
    \begin{split}
        \braket{\Phi^-_{\mathbf{k}_0',\bm{\beta}}|\Psi^+_{\mathbf{k}_,\bm{\alpha}}}&=\int dk_z'~\eta'(k_z')\sqrt{\frac{k'}{\mu}}\braket{\Psi^-_{\mathbf{k}',\bm{\beta}}|\Psi^+_{\mathbf{k},\bm{\alpha}}}=\int dk_z'~\eta'(k_z') \sqrt{\frac{k'}{\mu}}\delta(E'-E)S_{\bm{\beta},\bm{\alpha}}(E,\theta)\\
        &=\eta'(k_z')\sqrt{\frac{k'}{\mu}}\frac{\mu}{k_z'}S_{\bm{\beta},\bm{\alpha}}(E,\theta),
    \end{split}
    \label{eq:phi_psi_overlap}
\end{equation}
where $E'=k'^2+E_{\beta_1}^{(1)}+E_{\beta_2}^{(2)}$, $E=k^2+E_{\alpha_1}^{(1)}+E_{\alpha_2}^{(2)}$, and $\cos\theta=\hat{\mathbf{k}}'\cdot\hat{\mathbf{k}}$. Expanding $\ket{\Psi^+_{\mathbf{k},\bm{\alpha}}}$ in terms of $\ket{\Phi^+_{\mathbf{k}_0,\bm{\alpha}}}$ using Eq. (7) and rearranging yields Eq. (10) in the main text.

\section{Details on elastic two-body scattering in 2D}
\label{app:scattering_example}
Beginning from the Schr\"odinger equation for a central potential,
\begin{equation}
    \left[T+V(r)\right]\psi(\mathbf{r})=E\psi(\mathbf{r})
\end{equation}
and using the fact that the solution is separable
\begin{equation}
   \psi(\mathbf{r})= \psi(r,\theta)=\frac{u(r)}{\sqrt{r}}\frac{\cos(m\theta)}{\sqrt{\pi}},
\end{equation}
the radial Schr\"odinger equation is
\begin{equation}
   u''(r)+\left[k^2+\frac{1/4-m^2}{r^2}-\frac{2\mu}{\hbar^2}V(r)\right]u(r)=0,
\end{equation}
where $k^2=2\mu E/\hbar^2$ and $m$ is a non-negative integer. In two spatial dimensions, a plane wave incident along the $\hat{\mathbf{x}}$ axis can be expanded as 
\begin{equation}
    e^{ikx}=\sum_{m=0}^{\infty}\epsilon_m i^m\cos(m\theta)J_m(kr),
\end{equation}
where $J_m$ is the regular Bessel function of order $m$, and $\epsilon_m=2$ for $m>0$, $\epsilon_0=1$. The scattering wave function attains the asymptotic form
\begin{equation}
    \psi_k(r,\theta)\rightarrow \sum_{m=0}^{\infty}A_m (kr)^{-1/2}\cos(m\theta)\cos\left(kr-\frac{m\pi}{2}-\frac{\pi}{4}+\delta_m\right),
\end{equation}
where 
\begin{equation}
    A_m=2\epsilon_m i^m (2\pi)^{-1/2} e^{i\delta_m},
\end{equation}
and $\delta_m$ is the scattering phase shift. The scattering amplitude $f(E,\theta)$ can be computed from the phase shifts as \cite{10.1119/1.14623}
\begin{equation}
    f(E,\theta)=\sqrt{\frac{2}{\pi}}\sum_{m=0}^{\infty}\epsilon_m\cos(m\theta)e^{i\delta_m}\sin\delta_m
    \label{eq:famp_phases}
\end{equation}

By numerically integrating the potential using the variable phase method \cite{alma991072475789706532} (adapted to two spatial dimensions), one can obtain the partial-wave phase shifts $\delta_m$, which can then be combined to calculate the scattering amplitude, as in Eq. \eqref{eq:famp_phases}. The differential cross section can then be obtained as
\begin{equation}
    \frac{d\lambda}{d\theta}=\frac{1}{k}|f(E,\theta)|^2,
\end{equation}
and the integrated cross section can be obtained directly via the optical theorem,
\begin{equation}
    \lambda=\int d\theta\frac{d\lambda}{d\theta}=\frac{\sqrt{8\pi}}{k}\mathrm{Im}f(E,0).
\end{equation}  

In two spatial dimensions, the finite-volume $S$ matrix is related to the scattering amplitude by
\begin{equation}
    S(E,\theta)=\delta_{\theta}+\frac{\sqrt{2\pi}i}{kL}f(E,\theta),
\end{equation}
where $\delta_\theta=1$ if $\theta=0$ and $\delta_\theta=0$ if $\theta\neq 0$.

\section{Obtaining the overlap function from a quantum computer}
\label{app:overlap_quantum}
The overlap function, 
\begin{equation}
    \mathcal{C}_{\bm{\beta},\bm{\alpha}}(\mathbf{k}_0',\mathbf{k}_0;t)\equiv\braket{\Phi_{\mathbf{k}_0',\bm{\beta}}^-|e^{-iHt}|\Phi_{\mathbf{k}_0,\bm{\alpha}}^+},
\end{equation}
is the central quantity that we seek to obtain from quantum computation. Once the overlap function $\mathcal{C}(t)\equiv\mathcal{C}_{\bm{\beta},\bm{\alpha}}(\mathbf{k}_0',\mathbf{k}_0;t)$ is known at some number of snapshot times $t_i$, the subsequent Fourier transform that determines the scattering matrix is a trivial classical computation. As the initial state $\ket{\Phi_{\mathbf{k}_0,\bm{\alpha}}}$ and final state $\ket{\Phi_{\mathbf{k}_0',\bm{\beta}}}$ are in general distinct, their overlap is determined only up to an overall phase; that is,
\begin{equation}
    \braket{\Phi_{\mathbf{k}_0',\bm{\beta}}|\Phi_{\mathbf{k}_0,\alpha}}=|\braket{\Phi_{\mathbf{k}_0',\bm{\beta}}|\Phi_{\mathbf{k}_0,\alpha}}|e^{i\chi}\equiv N_0e^{i\chi},
\end{equation}
where $\chi$ does not have any physical meaning.  Expressing the overlap function as
\begin{equation}
    \mathcal{C}(t)=|\mathcal{C}(t)|e^{i\phi(t)+i\chi},
\end{equation}
we note that our formalism requires knowledge of both the time-dependent modulus $|\mathcal{C}(t)|$ and phase $\phi(t)$ in order to faithfully reconstruct the scattering amplitude via Fourier transform. 

\begin{figure}
    \centering
    \includegraphics[scale=0.9]{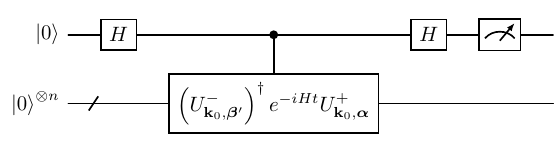}
    \caption{Hadamard test circuit that computes the real part of the overlap function, $\mathrm{Re}\braket{\Phi^-_{\mathbf{k}_0',\bm{\beta}}|e^{-iHt}|\Phi^+_{\mathbf{k}_0,\bm{\alpha}}}$, using a single set of particle registers.}
    \label{fig:overlap_circuit}
\end{figure}

Suppose that $U^+_{\mathbf{k}_0,\bm{\alpha}}$ $\left(U^-_{\mathbf{k}'_0,\bm{\beta}}\right)$ is a unitary operator that prepares the quantum state $\ket{\Phi_{\mathbf{k}_0,\bm{\alpha}}^+}$ $\left(\ket{\Phi_{\mathbf{k}_0',\bm{\alpha}}^-}\right)$ from the initially trivial configuration. We assume that each time we prepare these states, the relative phase $\chi$ is unchanged. Performing the Hadamard test illustrated in Fig. \ref{fig:overlap_circuit}, we can isolate the real part of the overlap function. After executing all circuit elements except for the measurement of the ancilla qubit, the state of the system is
\begin{equation}
\begin{split}
        \frac{1}{2}\left[\ket{0}^{\otimes n}+\left(U^-_{\mathbf{k}_0',\bm{\beta}}\right)^\dagger e^{-iHt}\ket{\Phi^+_{\mathbf{k}_0,\bm{\alpha}}}\right]\otimes \ket{0}
        + \frac{1}{2}\left[\ket{0}^{\otimes n}-\left(U^-_{\mathbf{k}_0',\bm{\beta}}\right)^\dagger e^{-iHt}\ket{\Phi^+_{\mathbf{k}_0,\bm{\alpha}}}\right]\otimes \ket{1},
        \end{split}
\end{equation}
and therefore the probability $P_0$ ($P_1$) of measuring the ancilla in state $\ket{0}$ ($\ket{1}$) is
\begin{equation}
\begin{split}
        P_0&=\frac{1}{2}+\frac{1}{2}\mathrm{Re}\left[\braket{\Phi^-_{\mathbf{k}_0',\bm{\beta}}|e^{-iHt}|\Phi^+_{\mathbf{k}_0,\bm{\alpha}}}\right],\\
        P_1&=\frac{1}{2}-\frac{1}{2}\mathrm{Re}\left[\braket{\Phi^-_{\mathbf{k}_0',\bm{\beta}}|e^{-iHt}|\Phi^+_{\mathbf{k}_0,\bm{\alpha}}}\right].
\end{split}
\end{equation}
Defining the random variable $A$ with probability mass function $P_A(x)\equiv P(A=x)$ given by
\begin{equation}
    P_A(1)=P_0,~~~P_A(-1)=P_1,
\end{equation}
the expectation value of $A$ yields the desired result,
\begin{equation}
\begin{split}
E[A]&=\mathrm{Re}\left[\braket{\Phi^-_{\mathbf{k}_0',\bm{\beta}}|e^{-iHt}|\Phi^+_{\mathbf{k}_0,\bm{\alpha}}}\right]\\
&=|\mathcal{C}(t)|\cos\left[\phi(t)+\chi\right].
\end{split}
\end{equation}
If the circuit in Fig. \ref{fig:overlap_circuit} is modified so that the first $H$ gate is immediately followed by an $S^\dagger$ gate, then the expectation value is modified to yield the imaginary part of the overlap function, 
\begin{equation}
\begin{split}
E[A]&=\mathrm{Im}\left[\braket{\Phi^-_{\mathbf{k}_0',\bm{\beta}}|e^{-iHt}|\Phi^+_{\mathbf{k}_0,\bm{\alpha}}}\right]\\
&=|\mathcal{C}(t)|\sin\left[\phi(t)+\chi\right].
\end{split}
\end{equation}
All observables depend on the quantity $|S|^2$, so that the phase $\chi$ is ultimately irrelevant. In the case of forward scattering, the initial and final states are equal, and therefore $\chi=0$. In this case, we can unambiguously extract both the modulus and phase of the $S$ matrix and thus utilize the optical theorem to obtain the total cross section.

The statistical error in the estimate provided by the Hadamard test is governed by the standard quantum limit; that is, $O(\epsilon^{-2})$ samples are required to achieve precision $\epsilon$. In principle, this can be improved to the Heisenberg limit $O(\epsilon^{-1})$ using quantum amplitude estimation \cite{Brassard:2000xvp} at the cost of increased circuit depth.

\section{Error analysis}
In this section, we discuss the various sources of error that arise when utilizing the time-dependent wave-packet formalism on a finite spatial lattice. We distinguish three classes of errors: (1) Discretization errors associated with the finite lattice spacing $a$; (2) errors from representing Gaussian wave packets on a finite lattice and propagating them to ``asymptotic separation''; and (3) errors related to sampling of the correlation function from a quantum computer and the subsequent Fourier transform to obtain the $S$ matrix. In principle there are other relevant sources of error, arising for example from Trotterization of the unitary time evolution operators or the approximation of arbitrary-angle single-qubit rotations by a finite sequence of Clifford + $T$ gates (for fault-tolerant implementations). A precise quantification of these errors will be highly dependent on the physical system under study, and we defer their consideration to future, more specialized works. 

\subsection{Discretization errors}

Discretization errors occur in any finite lattice calculation, leading to deviations from the corresponding continuum result. If the potential has a characteristic length scale $R_0$, then the size of these errors is governed by the dimensionless ratio $\chi =R_0/a$. Depending on the nature of the potential, the discretization error $\epsilon(\chi)$ can take many different parametric forms; for example, $\epsilon(\chi)\sim e^{-\chi^2}$ for Gaussian potentials and $\epsilon(\chi)\sim \chi^{-4}$ for Yukawa-like potentials. In the remainder of our error analysis, we will assume that the grid is chosen such that this discretization error is completely negligible compared to the other relevant errors. When this fails to be true, our error analysis can be considered as measuring the error relative to the underlying lattice calculation, rather than the corresponding continuum result. In principle, if the lattice spacing cannot feasibly be chosen small enough to suppress these discretization errors below some desired threshold, a series of calculations can be performed at different lattice spacings, and the continuum value can then be obtained by extrapolation to $a=0$.

\subsection{Wave-packet errors}
\label{app:wp_errors}
On a lattice with spacing $a$, the momentum resolution is limited by the Nyquist momentum $k_\mathrm{max}=\pi/a$. Thus, the wave packet that we represent should not contain significant contributions from momentum modes above this value; otherwise those modes will pollute via aliasing the lower-frequency modes of the lattice. Likewise, as discussed above in Sec. \ref{app:derivation}, in order for the energy projection to recover a unique value of the momentum $k_z$ from the wave packet, we must ensure that modes with $k_z<0$ are exponentially suppressed. Therefore we seek to minimize the extent of the wave packet that lies outside of the range $[0,k_\mathrm{max}]$. A straightforward calculation provides the following bound,
\begin{equation}
    1-\int_0^{k_\mathrm{max}}dk_z ~|\eta(k_z)|^2\leq \frac{1}{2\sqrt{\pi}\sigma}\left(\frac{1}{k_{0,z}}e^{-k_{0,z}^2\sigma^2}+\frac{1}{k_\mathrm{max}-k_{0,z}}e^{-(k_{0,z}-k_\mathrm{max})^2\sigma^2}\right),
\end{equation}
for a wave packet with momentum centered at $k_{0,z}$ and spatial width $\sigma$. When preparing for a practical calculation using the time-dependent wave-packet formalism, one will typically have a central momentum $k_{0,z}$ that is determined by the energies/momenta of interest. Thus, we proceed by assuming $k_{0,z}$ is an input parameter. Then for any $\delta>0$ and $k_{0,z}>0$ choosing 
\begin{equation}
    \sigma\geq \frac{1}{\sqrt{2}}\sqrt{W\left(\frac{2}{\pi\delta^2}\right)}\mathrm{max}[k_{0,z}^{-1},(k_{0,z}-k_\mathrm{max})^{-1}],
\end{equation}
where $W(x)$ is the Lambert-W (product-log) function, guarantees that 
\begin{equation}
    1-\int_0^{k_\mathrm{max}}dk_z~|\eta(k_z)|^2\leq \delta.
\end{equation}
To understand this bound more intuitively, we can use the fact that $W(x)\leq \log(x)$ for $x>e$. Then whenever $0<\delta<\sqrt{2/\pi e}$, it suffices to choose
\begin{equation}
    \sigma\geq \frac{1}{\sqrt{2}}\log^{1/2}\left(\frac{2}{\pi\delta^2}\right)\mathrm{max}[k_{0,z}^{-1},(k_{0,z}-k_\mathrm{max})^{-1}],
\end{equation}
to guarantee that the unwanted modes are suppressed below $\delta$. In other words, the error $\delta$ converges exponentially with increasing $\sigma$.

Next, we must guarantee that the wave packets can be propagated to ``asymptotic separation'' without incurring wrap-around errors from the periodic boundary conditions. Defining
\begin{equation}
    \ket{\Phi_{\mathbf{k}_0,\bm{\alpha}}(t)}\equiv e^{-iH_0t}\ket{\Phi_{\mathbf{k}_0,\bm{\alpha}}},
\end{equation}
and given some error threshold $\epsilon$, we seek $t_\infty$ such that
\begin{equation}
    \braket{H_\mathrm{int}}(t_\infty)\equiv\braket{\Phi_{\mathbf{k}_0,\bm{\alpha}}(t_\infty)|H_\mathrm{int}|\Phi_{\mathbf{k}_0,\bm{\alpha}}(t_\infty)}<\epsilon.
\end{equation}
The exact value of $\braket{H_\mathrm{int}}(t)$ is highly nontrivial to compute for a generic system, as it depends not only on the nature of the wave packets but also on the details of the interaction as well as the structure of the internal eigenstates $\ket{\psi_{\alpha_1}^{(1)}}$ and $\ket{\psi_{\alpha_2}^{(2)}}$. Nonetheless, we can obtain some estimate of the required value $t_\infty$ by neglecting the internal structure of the clusters and considering only the interaction of the relative wave packet through a Gaussian potential $V(r)=V_0e^{-r^2/2R_0^2}$, including the two nearest periodic images along the $z$ axis. The time-dependent expectation value of the interaction potential is then given by

\begin{equation}
    \braket{H_\mathrm{int}}(t)=\int d^3\vec{r}~V(r)|\braket{\phi_\mathrm{rel}(\mathbf{k}_0,\mathbf{r},t)}|^2,
\end{equation}
where 

\begin{equation}
\left|\braket{\phi_\mathrm{rel}(\mathbf{k}_0,\mathbf{r},t)}\right|^2\equiv\frac{1}{\sqrt{\pi}}\frac{1}{\sigma(t)} \mathrm{exp}\left[-\left(z-\frac{k_{0,z}}{\mu}t\right)^2/\sigma^2(t)\right],
\end{equation}
and
\begin{equation}
    \sigma(t)\equiv \sqrt{\sigma^2+\frac{t^2}{\mu^2}\frac{1}{\sigma^2}},
\end{equation}
is the width of the wave packet after propagation for time $t$ with the non-interacting Hamiltonian. 
Including the effect of the finite box length $L$ in the transverse directions (assuming 3 spatial dimensions), the expectation value of the potential evolves as
\begin{equation}
    \braket{H_\mathrm{int}}(t)=\frac{2\pi R_0^2}{L^2}\frac{V_0}{\sqrt{1+\sigma^2(t)/2R_0^2}}\left[\chi(-L,t)+\chi(0,t)+\chi(L,t)\right],
\end{equation}
where 
\begin{equation}
    \chi(z,t)\equiv \exp\left[-\frac{(z-k_{0,z}t/\mu)^2}{2R_0^2+\sigma^2(t)}\right].
\end{equation}
The time for the center of the wave-packet to propagate to $z=L/2$ is 
\begin{equation}
    t_\mathrm{mid}=\frac{\mu L}{2k_{0,z}},
\end{equation}
where the expectation value of the interaction is
\begin{equation}
    \braket{H_\mathrm{int}}(t_\mathrm{mid})=\frac{4\pi R_0^2}{L^2}\frac{V_0}{\sqrt{1+\sigma^2(t_\mathrm{mid})/2R_0^2}}\exp\left[-\frac{L^2}{8R_0^2+4\sigma^2(t_\mathrm{mid})}\right].
\end{equation}
For reasonable choices of $k_{0,z}$, $\sigma$, $L$, this will be close to --- though not exactly equal to --- the minimal value of $\braket{H_\mathrm{int}}(t)$. When $L/(\sigma k_{0,z}) \ll \sqrt{8R_0^2+4\sigma^2}$, then $\braket{H_\mathrm{int}}(t_\mathrm{mid})$ decreases exponentially with increasing $L$. For $L/(\sigma k_{0,z}) \gg \sqrt{8R_0^2+4\sigma^2}$, the exponential approaches a constant value $e^{-k_{0,z}^2\sigma^2}$, and the overlap decreases like $\braket{H_\mathrm{int}}(t_\mathrm{mid})\propto L^{-1/2}$. Although this result was derived by neglecting the internal structure of the scattering constituents as well as the details of the interaction, it should provide a useful heuristic for tuning the various parameters of the wave packet and lattice volume for a given simulation. In particular, adopting $t_\infty=t_\mathrm{mid}$ should be an appropriate starting point for further refinement.

\subsection{Numerical integration and sampling errors}

In Eq. (10) of the main text, the $S$ matrix is obtained by Fourier transforming the overlap function $\mathcal{C}_{\bm{\beta},\bm{\alpha}}(\mathbf{k}_0',\mathbf{k}_0;t)$ and multiplying by the dimensionless prefactor
\begin{equation}
    A_0(\mathbf{k}',\mathbf{k})\equiv\frac{1}{2\pi\eta'(k_z')\eta(k_z)}\frac{k_zk_z'}{\mu\sqrt{kk'}},
\end{equation}
where $\mathbf{k}=(k_{0,x},k_{0,y},k_z)$ and $\mathbf{k}'=(k_{0,x}',k_{0,y}',k_z')$. Thus, there are several sources of potential error:
\begin{enumerate}

    \item Error from truncating the range of integration 
    \begin{equation}
        \int_{-\infty}^{+\infty}dt~\mathcal{C}_{\bm{\beta},\bm{\alpha}}(\mathbf{k}_0',\mathbf{k}_0;t)e^{iEt}\rightarrow\int_{-t_0}^{+t_0}dt~\mathcal{C}_{\bm{\beta},\bm{\alpha}}(\mathbf{k}_0',\mathbf{k}_0;t)e^{iEt}
    \end{equation}
As described throughout this work, we can only realistically sample the correlation function within some time window $[-t_0,t_0]$. The truncation error then depends explicitly on the behavior of the correlation function at very long times. Assuming a short-ranged interaction, the Gaussian form of the relative wave packets suggests that the overlap function should damp exponentially at relatively short times compared to the dispersion time $t\ll 2\mu\sigma^2$. Due to the dispersive spreading of the wave packets, however, at very long times $t\gg 2\mu\sigma^2$ the correlation function can drop off no faster than $t^{-1/2}$. In practice, it is useful to introduce a window function when computing the Fourier transform of the correlation function. For example, we can adopt a (normalized) Gaussian filter function
\begin{equation}
    g(t)\equiv \frac{\sigma_E}{\sqrt{2\pi}} e^{-t^2\sigma_E^2/2},
\end{equation}
where $\sigma_E$ controls the width of the corresponding Gaussian in energy space. Letting $A_0=A_0(\mathbf{k}',\mathbf{k})$, $S(E)=S_{\bm{\beta},\bm{\alpha}}(E,\theta)$, and $\mathcal{C}(t)=\mathcal{C}_{\bm{\beta},\bm{\alpha}}(\mathbf{k}_0',\mathbf{k}_0;t)$, we find that introducing the filter function $g(t)$ corresponds to averaging the $S$ matrix over the equivalent energy profile,
\begin{equation}
    \tilde{S}(\sigma_E;E_0)\equiv A_0\int_{-\infty}^{+\infty}dt ~\mathcal{C}(t)e^{iE_0t}g(t)=\frac{A_0}{\sqrt{2\pi}\sigma_E}\int_{-\infty}^{+\infty} dE~ e^{-(E-E_0)^2/2\sigma_E^2}S(E).
\end{equation}
Thus we see directly that $\sigma_E$ controls the energy resolution $\Delta E\approx 2.355\sigma_E$ of the resulting energy-averaged $S$ matrix. The Gaussian window function $g(t)$ damps the correlation signal as $t\rightarrow \pm  \infty$, and it is straightforward to bound the residual error $\epsilon_\mathrm{tr}$ from truncating the range of the integral,
\begin{equation}
    \epsilon_\mathrm{tr}=\left|\int_{-\infty}^{-t_0}dt~\mathcal{C}(t)g(t)e^{iEt} +\int_{+t_0}^{+\infty}dt~\mathcal{C}(t)g(t)e^{iEt} \right|.
\end{equation}
It follows from unitarity of the $S$ matrix that $|\mathcal{C}(t)|\leq 1$ for all $-\infty<t<\infty$. The truncation error is then bounded by
\begin{equation}
   \epsilon_\mathrm{tr}\leq \sqrt{\frac{2}{\pi}}\frac{1}{t_0\sigma_E}e^{-t_0^2\sigma_E^2/2}.
\end{equation}
In practice, the resulting error $\epsilon_\mathrm{tr}$ will be significantly below this bound, as the correlation function $\mathcal{C}(t)$ is far from saturating the bound $|\mathcal{C}(t)|\leq 1$ for $|t|\geq t_0$. Assuming that the correlation function decreases monotonically for $|t|\geq t_0$, one can obtain a tighter bound on the truncation error 
\begin{equation}
    \epsilon_\mathrm{tr}\leq \sqrt{\frac{2}{\pi}}\frac{1}{t_0\sigma_E}e^{-t_0^2\sigma_E^2/2}\max\left[|\mathcal{C}(t_0)|,|\mathcal{C}(-t_0)|\right].
    \label{eq:tighter_bound}
\end{equation}
Although a priori one does not know the values of $|\mathcal{C}(\pm t_0)|$, Eq. \eqref{eq:tighter_bound} can be used together with the measured values of the correlation function in order to estimate the size of this error a posteriori.

    \item Error from discretizing the integral 
    \begin{equation}
        \int_{-t_0}^{t_0}dt~f(t)e^{iEt}\rightarrow \Delta t\sum_{t_i}~f(t_i)e^{iEt_i},
    \end{equation}
    where $f(t)\equiv \mathcal{C}(t)g(t)$. Suppose that the interval $[-t_0,t_0]$ is discretized among $N$ evenly spaced points with spacing $\Delta t=2t_0/N$. In order to resolve the phase factor at energy $E$, the grid spacing must satisfy $\Delta t<\pi/E$. Using a standard trapezoidal quadrature to perform the numerical integration, the resulting error is bounded by
    \begin{equation}
        \epsilon_\mathrm{int}\leq \frac{2t_0^3}{3N^2}M_1,
    \end{equation}
    where 
    \begin{equation}
        M_1\equiv \max_{t\in[-t_0,t_0]}|f''(t)+2iEf'(t)-E^2f(t)|.
    \end{equation}
    In practice, the size of this error can be estimated a posteriori by numerically approximating the first and second derivatives of $f(t)$ via finite differencing. A priori, it is guaranteed that the quadrature converges no worse than $O(1/N^2)$, assuming that the Nyquist criterion $\Delta t<\pi/E$ is satisfied.
    \item Errors resulting from the fact that the sampled values of the correlation function contain statistical noise. When obtained from sampling on the quantum computer, both the real and imaginary parts of $C(t_i)$ are subject to statistical error. Assuming that these quantities are estimated using a sample of $N_\mathrm{shot}$ Hadamard tests for each $t_i$, the standard deviation in the estimated value is
    \begin{equation}
        \sigma_{\mathrm{H},i}=\frac{\sqrt{1-x_i^2}}{\sqrt{N_\mathrm{shot}}},
    \end{equation}
    where $x_i=\mathrm{Re}[C(t_i)]$, $\mathrm{Im}[C(t_i)]$. We make the worst-case assumption that 
    \begin{equation}
        \sigma_{H,i}=\sigma_H\equiv \frac{1}{\sqrt{N_\mathrm{shot}}},
    \end{equation}
    for all $i =1,\ldots,N$. Letting $\hat{\mathcal{C}}(t_i)$ denote the value of $\mathcal{C}(t_i)$ estimated from $N_\mathrm{shot}$ samples each of the real and imaginary parts, we can model $\hat{\mathcal{C}}(t_i)$ as
    \begin{equation}
        \hat{\mathcal{C}}(t_i)=\mathcal{C}(t_i)+\eta_i+i\xi_i,
    \end{equation}
   where $\eta_i\sim\xi_i\sim\mathrm{iid}(0,\sigma_H^2)$ are independent, identically distributed (iid) random variables. Computing the variance of the trapezoidal quadrature of $\hat{\mathcal{C}}(t_i)g(t)e^{iEt}$ on $[-t_0,t_0]$ results in
    \begin{equation}
        \epsilon_\mathrm{stat}^2=\frac{4t_0\sigma^2}{N}T_N[g^2]-\frac{4t_0^2\sigma^2}{N^2}g(t_0)^2,
    \end{equation}
    where $T_N[g^2]$ is the trapezoidal estimate of $\int_{-t_0}^{t_0}dt~g^2(t)$. If the Gaussian smoothing function $g(t)$ is chosen so that $g(t_0)\approx 0$, then $T_N[g^2]\approx \sigma_E/(2\sqrt{\pi})$ and therefore
    \begin{equation}
        \epsilon_\mathrm{stat}\approx \sqrt{\frac{2\sigma_E t_0}{\sqrt{\pi}N}}\sigma_H=\sqrt{\frac{2\sigma_E t_0}{\sqrt{\pi}NN_\mathrm{shot}}}.
    \end{equation}
\end{enumerate}
The total error in the determination of the $S$ matrix is then bounded by
\begin{equation}
    \epsilon_\mathrm{tot}\leq A_0(\mathbf{k}',\mathbf{k})\left(\epsilon_\mathrm{tr}+\epsilon_\mathrm{int}+\epsilon_\mathrm{stat}\right).
\end{equation}

\bibliography{QScatter}